\documentclass[english,aps,pra,superscriptaddress,showpacs,twocolumn,floatfix,tocloft]{revtex4}

\usepackage{amsmath}
\usepackage{amssymb}
\usepackage{amsfonts}
\usepackage{bbm}
\usepackage{epsfig}
\usepackage{grffile}
\usepackage{babel}

\usepackage{enumitem} 

\usepackage[usenames,dvipsnames]{color}
\definecolor{grey}{rgb}{.5,.5,.5}
\definecolor{dblue}{rgb}{0,0,.5}
\definecolor{dgreen}{rgb}{0,.65,0}

\usepackage{bm}

\newcommand{\chem} {Department of Chemistry, School of Science and Research Center for Industries of the Future, Westlake University, Hangzhou, Zhejiang 310024, China}

\usepackage[colorlinks=true,citecolor=dblue,linkcolor=dblue]{hyperref}
\usepackage[all]{hypcap}
\usepackage{setspace}
\DeclareRobustCommand{\SkipTocEntry}[5]{}

\begin{document}

\title{From higher-order moments to time correlation functions in strongly correlated systems: A DMRG-based memory kernel coupling theory}

\date{September 2, 2025}

\author{Yunhao Liu}
\affiliation{\chem}

\author{Wenjie Dou}
\email{douwenjie@westlake.edu.cn} 
\affiliation{\chem}
\affiliation{Institute of Natural Sciences, Westlake Institute for Advanced Study, Hangzhou, Zhejiang 310024, China}
\affiliation{Key Laboratory for Quantum Materials of Zhejiang Province, Department of Physics, School of Science and Research Center for Industries of the Future, Westlake University, Hangzhou, Zhejiang 310024, China}

\begin{abstract}
We introduce a hybrid approach for computing dynamical observables in strongly correlated systems using higher-order moments. This method integrates memory kernel coupling theory (MKCT) with the density matrix renormalization group (DMRG), extending our recent work on MKCT [\href{https://arxiv.org/abs/2407.01923}{W. Liu, Y. Su, Y. Wang, and W. Dou, arXiv:2407.01923 (2024)}] to strongly correlated systems. The method establishes that correlation functions can be derived from the moments. Within our framework, operators and wavefunctions are represented as matrix product operators (MPOs) and matrix product states (MPSs), respectively. Crucially, the repeated application of the Liouville operator is achieved through an iterative procedure analogous to the DMRG algorithm itself. We demonstrate the effectiveness and efficiency of MKCT-DMRG by computing the spectral function of the Hubbard model. Furthermore, we successfully apply the method to compute the electronic friction in the Hubbard-Holstein model. In all cases, the results show excellent agreement with time-dependent DMRG (TD-DMRG) benchmarks. The advantage of MKCT-DMRG over TD-DMRG is the computational efficiency, which avoids expensive real-time propagation in TD-DMRG. These findings establish MKCT-DMRG as a promising and accurate framework for simulating challenging dynamical properties in strongly correlated quantum systems.
\end{abstract}


\maketitle

\section{Introduction}\label{sec:sec1}

The remarkable success of the density matrix renormalization group (DMRG)\cite{PhysRevLett.69.2863,PhysRevB.48.10345,SCHOLLWOCK201196} in approximating ground states of lattice systems stems fundamentally from its underlying matrix product state (MPS) representation. MPS provide an optimal description for gapped ground states in one dimension\cite{PhysRevB.73.094423,Hastings_2007} and serve as a powerful ansatz for gapless 1D systems, as well as ground states of finite-width two-dimensional\cite{annurev:/content/journals/10.1146/annurev-conmatphys-020911-125018} cylinders and strips.

MPS-based algorithms also enable efficient solutions to the time-dependent Schr\"odinger equation, collectively termed time-dependent DMRG (TD-DMRG)\cite{RevModPhys.77.259,PAECKEL2019167998}. These methods fall into two primary categories. The first category is based on approximating the formal time evolution operator $\mathrm{e}^{-\mathrm{i}H/\hbar t}$ or $\mathrm{e}^{-\mathrm{i}H/\hbar t}\left|\Psi\right\rangle$, e.g., Runge-Kutta\cite{PhysRevB.72.020404}, time-evolving block decimation (TEBD)\cite{PhysRevLett.91.147902,PhysRevLett.93.040502,PhysRevLett.98.070201,PhysRevLett.93.076401}, which globally propagates the wavefunction per time step via MPO/MPS multiplications. This temporarily increases the bond dimensions before compression to a target truncation threshold. In its original form, TEBD handles only nearest-neighbor interactions. A simple modification, exploiting swap gates\cite{Stoudenmire_2010} to move sites that are not next to each other to be temporarily adjacent, is effective for systems with only a modest number of beyond-nearest-neighbor interactions. The second category is the time-dependent variational principle (TDVP)\cite{Dirac_1930,PhysRevLett.107.070601,PhysRevB.94.165116}, which projects the Schr\"odinger equation onto the tangent space of fixed-bond-dimension MPS manifolds. This symplectic structure inherently conserves probability, energy, and other symmetry-protected integrals of motion\cite{PhysRevB.88.075133,PAECKEL2019167998}, while these quantities are not conserved in most other time-dependent MPS methods due to the necessity of truncation to keep an efficient MPS representation. It can also deal with finite long-range interactions, so it is easy to be used in treating the dynamics of real chemical systems\cite{10.1063/1.5125945,https://doi.org/10.1002/wcms.1614}. Therefore, it is one of the most attractive MPS time evolution algorithms. However, TD-DMRG faces challenges in long-time evolution due to error accumulation and entanglement growth, necessitating larger bond dimensions. 

Correlation functions can also be evaluated in the frequency domain via DMRG. Following the initial application of the Lanczos algorithm to zero-temperature spectral functions\cite{PhysRevLett.59.2999}, Lanczos DMRG\cite{PhysRevB.52.R9827,PhysRevB.85.205119} emerged by integrating DMRG with Lanczos recursion. This approach is computationally efficient and straightforward to implement, but limited to discrete spectra comprising only the lowest few excited states\cite{PhysRevB.60.335}—a consequence of compromised basis orthogonality due to MPS compression. Dynamical DMRG (DDMRG)\cite{PhysRevB.66.045114} achieves high accuracy through variational calculations in the frequency domain. However, its computational cost scales linearly with the number of target frequencies as each frequency point is evaluated independently. Similarly to the Lanczos algorithm, the Chebyshev polynomial expansion method\cite{RevModPhys.78.275}, recursively expands dynamical correlations, but does not demand the orthogonality of bases generated by the Chebyshev polynomial strictly. The Chebyshev MPS (CheMPS)\cite{PhysRevB.83.195115,doi:10.1021/acs.jpclett.1c02688} calculates correlation functions by combining the MPS with the Chebyshev polynomial expansion, offering an optimal accuracy-efficiency compromise. Although some of the finite-temperature extensions of frequency-domain algorithms exist, such as Lanczos DMRG with thermal state sampling\cite{PhysRevB.80.205117}, DDMRG and CheMPS using purification\cite{PhysRevLett.93.207204}-based thermal representations\cite{doi:10.1021/acs.jpclett.0c00905,PhysRevB.90.060406}, however, current frequency-domain finite-temperature DMRG methods face inherent trade-offs between accuracy and computational efficiency.

The projection operator formalism\cite{Nakajima1958projection, zwanzig1960projection, Zwanzig1961projection, mori1965projection,grabert2006projection} provides a powerful framework for evaluating correlation functions. The Mori–Zwanzig approach\cite{mori1965projection}, in particular, decomposes the system dynamics into a relevant subspace and its orthogonal complement using projection operators. This decomposition yields closed equations of motion for the relevant variables, while the orthogonal dynamics contribute as a memory kernel and stochastic noise. Crucially, the formalism itself does not prescribe a specific choice of relevant variables; these must be physically motivated for the system under study. The projection operator restricts the dynamics to the subspace spanned by the chosen relevant variables, with the orthogonal complement defined via an appropriate inner product (e.g., Kubo scalar product\cite{doi:10.1143/JPSJ.12.570} in quantum systems). This reformulates the correlation function computation into solving a generalized quantum master equation (GQME)\cite{10.1063/1.4948408,BHATTACHARYYA202416715} containing a memory kernel. The central challenge in practical applications lies in computing this memory kernel, which encapsulates the projected time evolution\cite{shi2004semiclassical, Kawasaki1973}.

In our previous work\cite{liu2024mkct}, we introduced an efficient approach for calculating time correlation functions, termed the memory kernel coupling theory (MKCT). MKCT defines a set of auxiliary high-order memory kernels governed solely by higher-order moments of the correlation functions, thus enabling the calculations of correlation functions without explicit time evolution. We subsequently developed a Pad\'e approximant-based truncation scheme for the kernel function\cite{10.1063/5.0273707}. However, both studies focused on impurity models.

Here, we present a novel approach, MKCT-DMRG, for calculating correlation functions in quasi-one-dimensional lattice models with strong correlations. This method synergistically integrates MKCT with DMRG, extending our recent work on MKCT to strongly correlated systems. We demonstrate MKCT-DMRG's efficacy by computing correlation functions for the Hubbard model. Furthermore, we apply it to calculate electronic friction\cite{PhysRevB.96.104305,PhysRevB.97.064303} in the Hubbard-Holstein model. In both cases, the results show close agreement with the TD-DMRG benchmarks. These findings establish MKCT-DMRG as an accurate framework for simulating challenging dynamical properties in strongly correlated quantum systems.

The paper is structured as follows. Sec. \ref{sec:sec2A} reviews the MKCT framework; Sec. \ref{sec:sec2B} introduces the MKCT-DMRG approach. In Sec. \ref{sec:sec3}, we apply MKCT-DMRG: Sec. \ref{sec:sec3A} presents calculations of correlation functions for the Hubbard model, and Sec. \ref{sec:sec3B} details the electronic friction calculations for the Hubbard-Holstein model. We conclude in Sec. \ref{sec:sec4}.

\section{Theory and Methodology} \label{sec:sec2}

\subsection{Memory kernel coupling theory}\label{sec:sec2A}

The aim of developing MKCT is to calculate the following correlation function,
\begin{equation}\label{eqn-1}
    C_{AB}\left(t\right)\equiv \left\langle A(t)B(0)\right\rangle = \operatorname{tr}\left(\mathrm{e}^{\mathrm{i}H/\hbar t}A \mathrm{e}^{-\mathrm{i}H/\hbar t} B \rho\right),
\end{equation}
where $A$ and $B$ are arbitrary operators and $\rho$ is the initial density matrix.

According to the Mori-Zwanzig formalism\cite{Nakajima1958projection,zwanzig1960projection,mori1965projection}, the derivative of operator $A$ in the Heisenberg picture satisfies 
\begin{equation}\label{eqn-2}
    \dot{A}(t)=\Omega A(t)+\int_{0}^{t} d \tau K(\tau) A(t-\tau)+f(t).
\end{equation}
We define the higher-order moments
\begin{equation}\label{eqn-3}
    \Omega_n \equiv \frac{ ( \mathcal{L}^n A, B)} {(A, B)},
\end{equation}
and the auxiliary kernels
\begin{equation}\label{eqn-4}
    K_n(t) \equiv \frac{(\mathcal{L}^n f(t), B)} {(A, B)},
\end{equation}
where $\mathcal{L}(\cdot) \equiv \mathrm{i}\left[H, \cdot\right]/\hbar$ is the Liouville operator, and we omit the subscript “1” of the first-order moment $\Omega_1$ and auxiliary kernel $K_1(t)$. $f(t)$ in Eq.~(\ref{eqn-2}) is referred to as the random fluctuation operator, defined as
\begin{equation}\label{eqn-5}
    f(t)\equiv \mathrm{e}^{\bm{\mathcal{Q}} \mathcal{L} t } \bm{\mathcal{Q}} \mathcal{L}A,
\end{equation}
where we define the projection operator 
\begin{equation}\label{eqn-6}
    \bm{\mathcal{P}} (\cdot) \equiv \frac{(\cdot, B)} {(A, B)} A,
\end{equation}
and the complementary projection operator is $\bm{\mathcal{Q}} \equiv \bm{1} - \bm{\mathcal{P}}$.

Combining Eq. (\ref{eqn-1}) and Eq. (\ref{eqn-2}), we derive the GQME\cite{10.1063/1.4948408,BHATTACHARYYA202416715}, which is the derivative of Eq. (\ref{eqn-1}),
\begin{equation}\label{eqn-7}
    \dot{C}_{AB}(t)=\Omega {C}_{AB}(t)+\int_{0}^{t} d \tau K(\tau) {C}_{AB}(t-\tau)+F(t),
\end{equation}
where the inhomogeneous term $F(t)\equiv \left\langle f(t)B\right\rangle$ can be removed via the proper choice of projection operator\cite{liu2024mkct}.

Although several works\cite{kelly_generalized_2016, montoya-castillo_approximate_2016, yan_theoretical_2019} have attempted to approximate the memory kernel function $K(t)$, obtaining it with high accuracy remains challenging. The key aspect of our previous work lies in extending the definitions of $\Omega$ and $K(t)$ within the Mori GQME framework\cite{liu2024mkct} and introducing higher-order moments of the correlation functions.  Subsequently, we have introduced Pad\'e approximant to approximate the kernel function $K(t)$\cite{10.1063/5.0273707}.

We have shown that the higher-order kernels satisfy the following coupled ordinary differential equations (ODEs)\cite{liu2024mkct}: 
\begin{equation}\label{eqn-8}
    \dot{K}_n(t) = K_{n+1}(t) - \Omega_n K(t), 
\end{equation}
with initial conditions $K_{n}(0) = \Omega_{n+1} - \Omega_n \Omega_1$. This result is compelling as it reveals that the quantum dynamics can be fully encoded in a system of interconnected ODEs for the higher-order kernels. Notably, the only parameters needed to integrate $\{K_n(t)\}$ are $\{\Omega_n\}$, meaning that our MKCT framework eliminates the time evolution of operator or wavefunction directly. 

To fit the $n$-th order kernel $K_n(t)$, Ref.\cite{10.1063/5.0273707} introduces the following Pad\'{e} approximant, 
\begin{equation}\label{eqn-9}
    K_n(t) \approx \frac{p_{M_1}(t)}{q_{M_2}(t)} = \frac{\sum_{j=0}^{M_1}a_j t^{j}}{1 + \sum_{j=1}^{M_2}b_j t^{j}},
\end{equation}
where $ p_{M_1}(t)$ and $q_{M_2}(t)$ are polynomials of orders $M_1$ and $M_2$, respectively. The coefficients $\{a_i\}$ and $\{b_i\}$ are calculated using the Python library SciPy, which implements the standard Pad\'{e} approximant procedure as described in Ref.~\cite{Baker1996pade}. Overall, Eq.~(\ref{eqn-9}) provides a numerically stable truncation for the MKCT Eq.~(\ref{eqn-8}). We next demonstrate that the coefficients of Eq.~(\ref{eqn-9}) can be evaluated with higher-order moments $\{ \Omega_n \}$.

To begin with, notice that the $m$-th derivative of kernel $K_n(t)$ evaluated at $t=0$ is 
\begin{equation}\label{eqn-10}
    K_n^{(m)}(0) = \frac{(\mathcal{L}^n (\bm{\mathcal{Q}} \mathcal{L})^{m+1} A, B)} {(A, B)}.
\end{equation}
By expanding the operator $\bm{\mathcal{Q}}$, we derive the following recursion relation
\begin{equation}\label{eqn-11}
    K_n^{(m)} = K_{n+1}^{(m-1)} - \Omega_n \tilde{\Omega}_m,
\end{equation}
where we introduce the auxiliary moment 
\begin{equation}\label{eqn-12}
    \tilde{\Omega}_m \equiv \frac{(\mathcal{L} (\bm{\mathcal{Q}} \mathcal{L})^{m} A, B)} {(A, B)}.
\end{equation}
Recursively applying Eq.~(\ref{eqn-12}) eventually leads the following expression: 
\begin{equation}\label{eqn-13}
    K_{n}^{(m)}(0) = \Omega_{m+n+1} - \sum_{j=0}^{m} \Omega_{n+j} \tilde{\Omega}_{m-j}.
\end{equation}
Similarly, we can derive the recursion relation of the auxiliary moments,
\begin{equation}\label{eqn-14} 
\tilde{\Omega}_m =\Omega_{m+1} - \sum_{j=1}^{m}\Omega_{j}\tilde{\Omega}_{m-j}, 
\end{equation}
which means that the auxiliary moments $\{\tilde{\Omega}_m\}$ can be readily obtained by the moments $\{\Omega_n\}$. 

Substituting Eq. (\ref{eqn-14}) to Eq. (\ref{eqn-13}), we can derive the $m$-th derivative of the kernel $K_n(t)$ at $t=0$, that is, $K_n^{(m)}(0)$, which will be used to fit the coefficients of Pad\'e approximant (Eq.~(\ref{eqn-9})). Therefore, the key of our approach is how to evaluate the moments $\{\Omega_n\}$.

\subsection{Calculating moments by DMRG}\label{sec:sec2B}

Here, we propose an algorithm to extend MKCT to quasi-one-dimensional strongly correlated systems by combining MKCT and DMRG. In DMRG algorithms, the wavefunction and operator could be represented as matrix product state (MPS) and matrix product operator (MPO), respectively,
\begin{align}
|\Psi\rangle =& \sum_{\{\alpha_i\},\{\sigma_i\}} A_{\alpha_{1}}^{\sigma_{1}} A_{\alpha_{1} \alpha_{2}}^{\sigma_{2}} \cdots A_{\alpha_{N-1}}^{\sigma_{N}}\left|\sigma_{1} \sigma_{2} \cdots \sigma_{N}\right\rangle,  \label{eqn-15} \\
O =& \sum_{\{\omega_i\},\{\sigma_i\},\left\{\sigma_i^{\prime}\right\}} W_{\omega_{1}}^{\sigma_{1}^{\prime}, \sigma_{1}} W_{\omega_{1} \omega_{2}}^{\sigma_{2}^{\prime}, \sigma_{2}} \cdots W_{\omega_{N-1}}^{\sigma_{N}^{\prime}, \sigma_{N}}  \nonumber \\
& \left|\sigma_{1}^{\prime} \sigma_{2}^{\prime} \cdots \sigma_{N}^{\prime}\right\rangle\left\langle\sigma_{N} \sigma_{N-1} \cdots \sigma_{1}\right|,
\end{align}
where $\sigma_i$ and $\alpha_i$ (or $\omega_i$) in tensor $A_{\alpha_{i-1} \alpha_{i}}^{\sigma_{i}}$ (or $W_{\omega_{i-1} \omega_{i}}^{\sigma_{i}^{\prime},\sigma{i}}$) are the indices of physical bond and virtual bond, respectively. Eq.~(\ref{eqn-15}) can be rebuilt as a mixed-canonical MPS,
\begin{equation}
    |\Psi\rangle =\sum_{\sigma_i;\alpha_{i-1},\alpha_i} M_{\alpha_{i-1},\alpha_i}^{\sigma_i}\left|\mathcal{L}_{\alpha_{i-1}}^{[1:i-1]}\right\rangle \left|\sigma_i\right\rangle \left|\mathcal{R}_{\alpha_i}^{[i+1:n]}\right\rangle,
\end{equation}
where $\left|\mathcal{L}_{\alpha_{i-1}}^{[1:i-1]}\right\rangle$ and $\left|\mathcal{R}_{\alpha_i}^{[i+1:n]}\right\rangle$ are block configurations with the left and right orthonormal basis, respectively, and the site $i$ with arbitrary tensor component $M_{\alpha_{i-1},\alpha_i}$ is called active site or orthogonality center. For introducing the TDVP algorithm, we also define the degrees of freedom $C_{\alpha_{i-1},\alpha_i}$ on the virtual bond between the sites $i$ and $i+1$ by the following formula:
\begin{equation}
    |\Psi\rangle =\sum_{\alpha_l,\alpha_r} C_{\alpha_l,\alpha_r}\left|\mathcal{L}_{\alpha_l}^{[1:i]}\right\rangle  \left|\mathcal{R}_{\alpha_r}^{[i+1:n]}\right\rangle.
\end{equation}

To calculate the moments by Eq.~(\ref{eqn-3}), the initial wavefunction at equilibrium must be obtained. Our implementation utilizes the purification approach\cite{PhysRevLett.93.207204} for thermal state representation, where mixed states are encoded as pure states in an enlarged Hilbert space P$\otimes$Q formed by adding an auxiliary space Q to the physical space P. Thermal equilibrium states at specific temperature $\left|\psi_\beta\right\rangle$ are generated through imaginary-time evolution from the maximally entangled identity state:
\begin{align}
\left|\psi_\beta\right\rangle&=\frac{1}{\sqrt{Z}}\mathrm{e}^{-\beta H}\left|I\right\rangle,   \\
\left|I\right\rangle&\equiv\sum_n\left|n,\tilde{n}\right\rangle, 
\end{align}
where $\left|\tilde{n}\right\rangle$ is the state in auxiliary space that is same as the state $\left|n\right\rangle$ in physical space.

For time evolution operators $U(\tau)=\mathrm{e}^{-\tau H}$, we adopt a hybrid strategy leveraging both TEBD and TDVP methods. For any nearest-neighbor Hamiltonian with $2N$ sites
\begin{equation}
    H=\sum_{j=1}^{2N-1} h_{j,j+1},
\end{equation}
where $h_{j,j+1}$ acts on the $j$-th site and $(j+1)$-th site. The Hamiltonian can be decomposed into two parts 
\begin{align}
    H &= H_1 + H_2,  \\
    H_1 &= \sum_{j=1}^{N} h_{2j-1,2j},   \\
    H_2 &= \sum_{j=1}^{N-1} h_{2j,2j+1}.  
\end{align}
The TEBD algorithm employs a second-order Suzuki-Trotter decomposition:
\begin{equation}
\mathrm{e}^{-\tau H}=\mathrm{e}^{-\tau H_{1}/2} \mathrm{e}^{-\tau H_{2}} \mathrm{e}^{-\tau H_{1}/2}+\mathcal{O}\left(\tau^{3}\right).  
\end{equation}
In imaginary time evolution, a larger time step is typically necessitated at lower temperatures. This poses a challenge for the TEBD algorithm, which exhibits pronounced numerical instability with large steps. Consequently, the TDVP method is utilized for the time evolution in the low-temperature regime. 

The TDVP-based propagation governed by the variational condition:
\begin{equation}
\min \left \| H\left|\psi(t)\right\rangle -\mathrm{i}\hbar\frac{\partial}{\partial t} \left|\psi(t)\right\rangle \right \|.
\end{equation}
With the MPS framework, this can be achieved by projecting $H|\psi(t)\rangle$ onto the tangent space of the given $|\psi(t)\rangle$ in the tensor manifold. The projector on the tangent space is defined as
\begin{equation}\label{eqn-proj}
    P_{T,|\psi(t)\rangle}=\sum_{i=1}^{n} P_{i-1}^{L} \otimes I_{i} \otimes P_{i+1}^{R}-\sum_{i=1}^{n-1} P_{i}^{L} \otimes P_{i+1}^{R},
\end{equation}
where $P_i^L$ and $P_i^R$ are the left and right block projectors, respectively,
\begin{subequations}
\begin{align}
    P_{i}^{L}&=\sum_{\alpha_{i}}\left|\mathcal{L}_{\alpha_{i}}^{[1: i]}\right\rangle\left\langle\mathcal{L}_{\alpha_{i}}^{[1: i]}\right|, \\
    P_{i}^{R}&=\sum_{\alpha_{i}}\left|\mathcal{R}_{\alpha_{i-1}}^{[i: n]}\right\rangle\left\langle\mathcal{R}_{\alpha_{i-1}}^{[i: n]}\right| .
\end{align}
\end{subequations}
By inserting the projector (Eq.~(\ref{eqn-proj})) to the Schr\"odinger equation $\mathrm{i}\hbar\partial_t|\psi(t)\rangle=H|\psi(t)\rangle$, we derive its local version,
\begin{equation}
    \mathrm{i}\hbar\frac{\partial}{\partial t}|\psi(t)\rangle=P_{T,|\psi(t)\rangle}H|\psi(t)\rangle.
\end{equation}
The above equation can be solved approximately by solving $n$ forward-evolving equations and $n-1$ backward-evolving equations, respectively,
\begin{subequations}
    \begin{align}
    \mathrm{i}\hbar \frac{\partial}{\partial t}|\psi(t)\rangle &=\sum_{i=1}^{n} P_{i-1}^{L} \otimes \mathbf{1}_{i} \otimes P_{i+1}^{R} H|\psi(t)\rangle,  \label{eqn-forward} \\
    \mathrm{i}\hbar \frac{\partial}{\partial t}|\psi(t)\rangle &=-\sum_{i=1}^{n-1} P_{i}^{L} \otimes P_{i+1}^{R} H|\psi(t)\rangle .   \label{eqn-backward} 
    \end{align}
\end{subequations}
As a result, Eqs. (\ref{eqn-forward}) and (\ref{eqn-backward}) can work in the sequence defined by the order of the local sites, instead of having to work with the full MPS $|\psi\rangle$. Therefore, in practice, the $i$-th term of Eqs. (\ref{eqn-forward}) and (\ref{eqn-backward}) can be integrated with a time-differential equation as follows:
\begin{subequations}
    \begin{align}
    \mathrm{i}\hbar\dot{\mathbf{M}}^{\sigma_i}&= \sum_{\sigma_i^\prime}H_{\operatorname{eff},1}^{\sigma_i\sigma_i^\prime}\mathbf{M}^{\sigma_i^\prime},  \label{eqn-for-mat} \\
    \mathrm{i}\hbar\dot{\mathbf{C}^i}&= -H_{\operatorname{eff},0}^{i}\mathbf{C}^i,  \label{eqn-back-mat}
    \end{align}
\end{subequations}
where $\mathbf{M}^{\sigma_i}\equiv \{M_{\alpha_{i-1}\alpha_i}^{\sigma_i}\}$, $\mathbf{C}^{i}\equiv \{C_{\alpha\beta}\}$ are the matrix representations of the corresponding tensor; $H_{\mathrm{eff}, 1}^{\sigma_{i}, \sigma_{i}^{\prime}}  =\left\langle\sigma_{i}\right| H_{\mathrm{eff}, 1}^{i}\left|\sigma_{i}^{\prime}\right\rangle$, in which $H_{\mathrm{eff}, 1}^i$ is the one-site effective Hamiltonian, and $H_{\mathrm{eff}, 0}^i$ in Eq.~(\ref{eqn-back-mat}) is the zero-site effective Hamiltonian, defined as follows, respectively,
\begin{subequations}
    \begin{align}
    H_{\mathrm{eff}, 1}^{i} & =\left\langle\mathcal{R}_{\alpha_{i}}^{[i+1: n]}\right| \otimes\left\langle\mathcal{L}_{\alpha_{i-1}}^{[1: i-1]}\right| H\left|\mathcal{L}_{\alpha_{i-1}^{\prime}}^{[1: i-1]}\right\rangle \otimes\left|\mathcal{R}_{\alpha_{i}^{\prime}}^{[i+1: n]}\right\rangle,   \\
    H_{\mathrm{eff}, 0}^{i} & =\left\langle\mathcal{R}_{\alpha_{i, r}}^{[i+1: n]}\right| \otimes\left\langle\mathcal{L}_{\alpha_{i, l}}^{[1: i]}\right| H\left|\mathcal{L}_{\alpha_{i, l}^{\prime}}^{[1: i]}\right\rangle \otimes\left|\mathcal{R}_{\alpha_{i, r}^{\prime}}^{[i+1: n]}\right\rangle.
    \end{align}
\end{subequations}

The above is the single-site TDVP (1TDVP), which maintains fixed bond dimensions during the time evolution. The TDVP framework also offers a two-site approach, that is, 2TDVP, in which the two-site and one-site effective Hamiltonian are applied to Eqs.~(\ref{eqn-for-mat}) and (\ref{eqn-back-mat}), respectively. The bond dimensions of MPS in 2TDVP can be adaptively increased, but the computational costs also increase. The TDVP algorithm suffers from a non-negligible projection error onto the manifold of MPS at given small bond dimension. Moreover, time evolution of MPS will suffer from a substantial projection error. Therefore, it is necessary to first perform time evolution using other methods to increase bond dimensions for a few steps before performing time evolution using TDVP. We summarize our strategy for obtaining the thermal equilibrium state according to DMRG as follows\cite{PhysRevX.11.031007}:
\begin{enumerate}[noitemsep, topsep=0pt, partopsep=0pt, parsep=0pt, itemindent=*, leftmargin=*]
    \item Apply the TEBD with high accuracy up to a time $\tau_\text{TEBD}$ to obtain a MPS with a suitably large bond dimension.
    \item Employ 2TDVP to further increase the bond dimension.
    \item Upon reaching maximum bond dimension, switch to 1TDVP for computational efficiency.
\end{enumerate}

\begin{figure*}[htbp]
    \centering
    \includegraphics[width=0.75\linewidth]{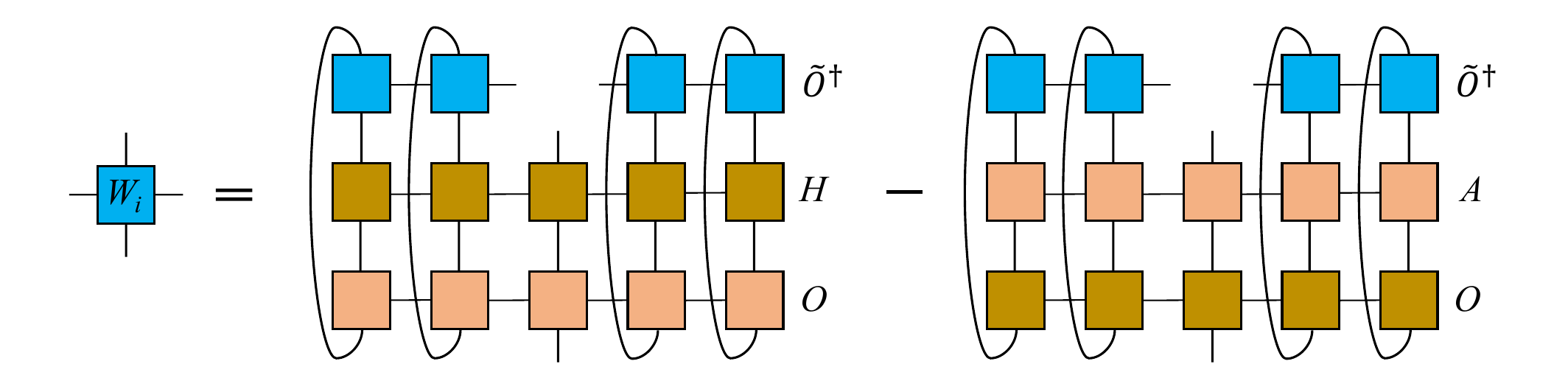}
    \caption{\label{fig-1} Diagram expression of (one-site) variational optimization of local site $W_i$ in MPO according to Eq.~(\ref{eqn-mkct-dmrg}).} 
\end{figure*}

After obtaining the thermal equilibrium state, for DMRG algorithms, the operation of the inner product of operators $A$ and $B$ is relatively easy to realize; however, the most crucial issue is how to perform the operation $\mathcal{L}O=\mathrm{i}\left[H,O\right]/\hbar$, where $O$ is an arbitrary operator. 

Generally, we should avoid performing MPO multiplication as soon as possible to prevent the rapid growth of virtual bond dimensions. To obtain the result of operation $HO-OH$, we initialize a random MPO $\tilde{O}$ to approximate $HO-OH$ by variational minimization the functional
\begin{align}
\mathcal{F} &\equiv \operatorname{Tr} \left \|\tilde{O} - \left(HO-OH\right)\right\|^2 \nonumber \\
&= \operatorname{Tr} \left(\tilde{O} - \left(HO-OH\right) \right)^\dagger \left(\tilde{O} - \left(HO-OH\right) \right).
\end{align}

Based on the MPO representation, $\tilde{O}$ can be obtain by solving the uncoupled linear equations $\partial \mathcal{F} / \partial{W_i}= 0$ locally by sweeps following the philosophy of DMRG, 
\begin{equation}\label{eqn-mkct-dmrg}
    \frac{\partial \mathcal{F}}{\partial W_i}= \operatorname{Tr} \left[\frac{\partial \tilde{O}^\dagger}{\partial W_i}\tilde{O} - \frac{\partial \tilde{O}^\dagger}{\partial W_i}\left(HO-OH\right) \right] =0,
\end{equation}
where $W_i\equiv W_{\omega_{i-1} \omega_{i}}^{\sigma_{i}^{\prime},\sigma{i}}$. The process of solving the above equation iteratively is illustrated in Fig. \ref{fig-1}.

The code implementing the MKCT presented in Sec. \ref{sec:sec2A} are available from the Zenodo repository\cite{bi_2025_15393237}. We developed a DMRG module building upon tensor contraction engine and MPS/MPO infrastructure in ITensor library\cite{itensor} to calculate the moments (Eq. (\ref{eqn-3})) by the method presented in Sec. \ref{sec:sec2B}. All our calculations in Sec. \ref{sec:sec3} were performed by combining these two parts of code. All TD-DMRG calculations were performed using the ITensor library\cite{itensor}, leveraging its native implementations of TEBD and TDVP algorithms. To enable finite-temperature simulations, we extended the ITensor framework through purification method.

\section{Results and Discussions} \label{sec:sec3}

In the following sections, we employ our MKCT-DMRG approach to assess the time-dependent properties of 1D strongly correlated systems. As a benchmark, we utilize the TD-DMRG method\cite{RevModPhys.77.259,PAECKEL2019167998}. During the real-time evolution, we utilize the one-site TDVP (1TDVP) method\cite{Dirac_1930,PhysRevLett.107.070601} to assess time correlation functions. The initial finite-temperature thermal states\cite{doi:10.1021/acs.jpclett.0c00905,PhysRevB.90.060406}, prepared via the purification method\cite{PhysRevLett.93.207204}, are obtained by imaginary-time evolution. 

\subsection{Tests of correlation functions} \label{sec:sec3A}

\begin{figure*}[htbp]
    \centering
    \includegraphics[width=0.85\linewidth]{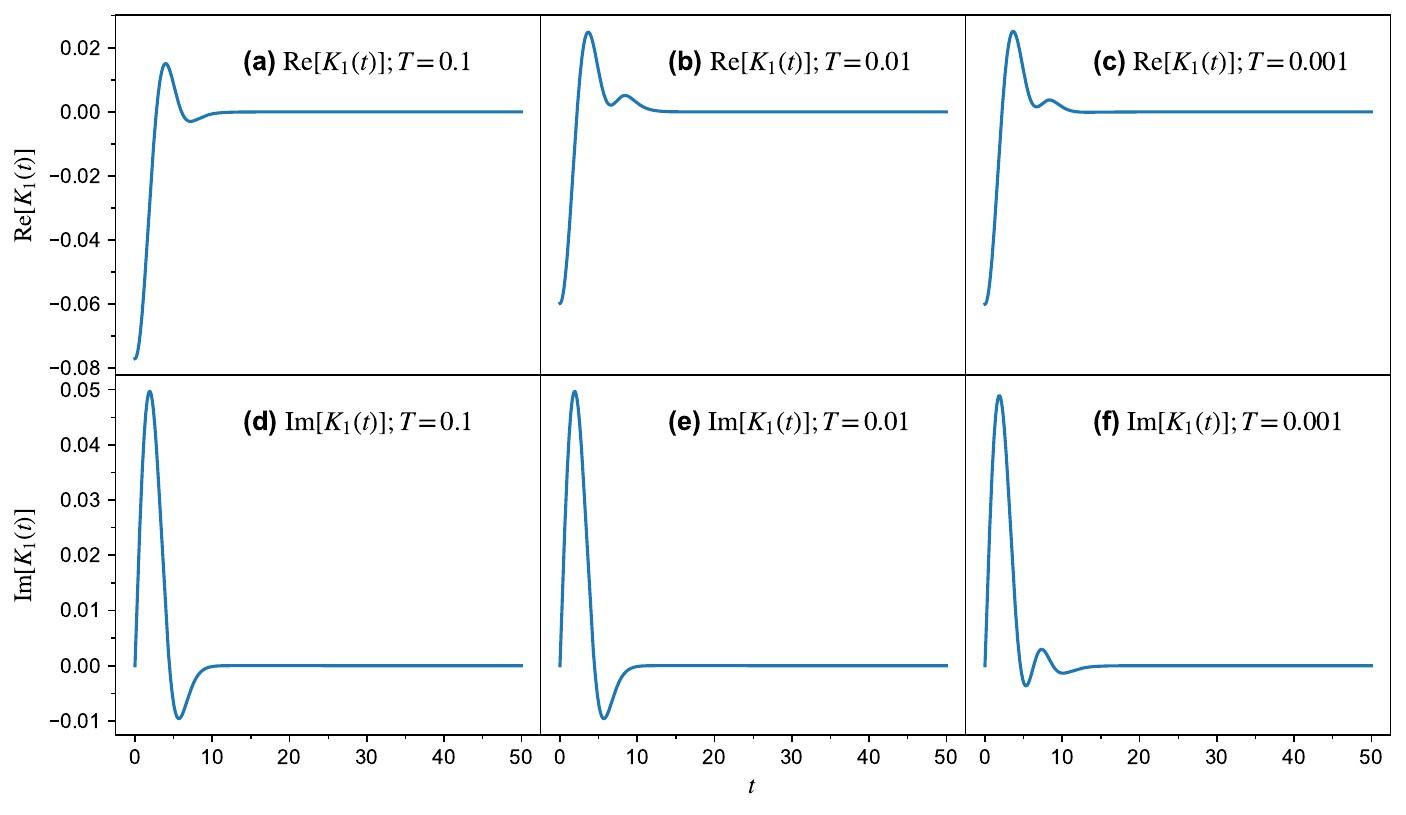}
    \caption{\label{fig-2}{Memory kernels $K_1(t)$ used to obtain correlation function $C(t)$ of Hubbard model. (a)-(c) the real parts of memory kernels ($\operatorname{Re}\left[K_1(t)\right]$) with diverse temperature; (d)-(f) the imaginary parts of memory kernels ($\operatorname{Im}\left[K_1(t)\right]$) with diverse temperature. The Pad\'e orders are $[M_1/M_2]=[4/15],[9/20],[9/20]$ for temperature $T=0.1,0.01,0.001$, respectively.}}
\end{figure*}

\begin{figure*}[htbp]
    \centering
    \includegraphics[width=0.85\linewidth]{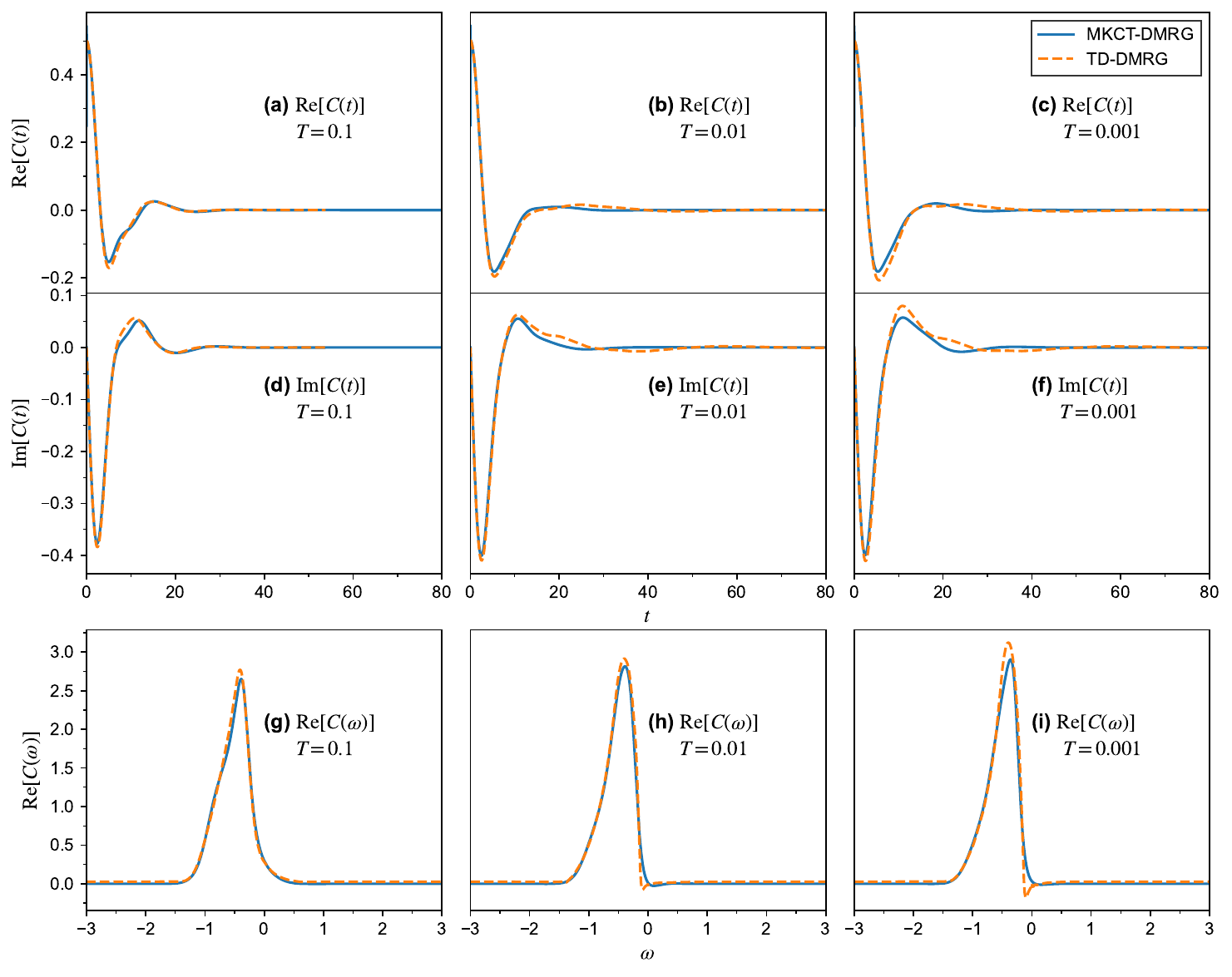}
    \caption{\label{fig-3}{Time-domain correlation functions $C(t)$ and frequency-domain correlation functions $C(\omega)$ of Hubbard model. (a)-(c) the real parts of correlation functions in time domain ($\operatorname{Re}\left[C(t)\right]$) with diverse temperature; (d)-(f) the imaginary parts of correlations functions in time domain ($\operatorname{Im}\left[C(t)\right]$) with diverse temperature; (g)-(i) the real parts of correlation functions in frequency domain ($\operatorname{Re}\left[C(\omega)\right]$) with diverse temperature. The Pad\'e orders are $[M_1/M_2]=[4/15],[9/20],[9/20]$ for temperature $T=0.1,0.01,0.001$, respectively.}}
\end{figure*}

The Hubbard model provides an ideal testbed for studying strongly correlated systems. In our calculations, the model is written as
\begin{equation}
    H_\text{Hub}=\epsilon \sum_{i,\sigma} n_{i,\sigma} + t\sum_{i,\sigma}\left( c_{i,\sigma}^\dagger c_{i+1,\sigma}+ \text{h.c.} \right) + U\sum_i n_{i\uparrow }n_{i\downarrow },
\end{equation}
where $c^\dagger_{i\sigma}$,$c_{i\sigma}$ are fermionic creation and annihilation operators of spin $\sigma$ on site $i$, and $n_{i\sigma}\equiv c_{i\sigma}^\dagger c_{i\sigma}$. To test our MKCT-DMRG approach, we will now calculate the following time correlation function,
\begin{equation}\label{eqn-corr-ca}
    C_{i,j}(t)=\left\langle c_i^\dagger(t) c_j \right\rangle,
\end{equation}
where $c_i^\dagger(t)=\mathrm{e}^{\mathrm{i}H/\hbar t}c_i^\dagger\mathrm{e}^{\mathrm{i}H/\hbar t}$.
Performing a Fourier transform on the above equation, that is, $C_{i,j}(\omega)=\int\left\langle c_i^\dagger(t) c_j \right\rangle\mathrm{e}^{\mathrm{i}\omega t}\mathrm{d}\omega$, we can obtain the photoelectron spectrum, defined as $\frac{1}{\pi}\operatorname{Re}\left[C_{ij}(\omega)\right]$. In our calculations, we set $i=j=1$, and we omit the subscript "$i$" and "$j$" in the following. In this section, the parameters of Hubbard model are $\epsilon=-0.5, t = 0.3, U = 1.0$, and we set $\hbar=k_B=1$ in our calculations. The number of sites is 100.

Fig.~\ref{fig-2} presents $K_1(t)$ with temperature $T=0.1, 0.01$ and $0.001$ calculated by Eq.~(\ref{eqn-9}), which is used to calculate the correlation function $C(t)$ by Eq.~(\ref{eqn-7}). Compared with the time correlation functions in Fig.~\ref{fig-3}, it shows that the kernels $K_1(t)$ decay to zero faster than the time correlation functions $C(t)$, demonstrating that the kernels capture the dynamics on a shorter timescale. 

We compare the the time-domain and frequency-domain correlation functions ($C(t)$ and $C(\omega)$) calculated from MKCT-DMRG versus the results calculated from TD-DMRG at temperature $T=0.1, 0.01$ and $0.001$ in Fig.~\ref{fig-3}. All results calculated by MKCT-DMRG and TD-DMRG are consistent very well. Notice that the Pad\'e orders $[M_1/M_2]$ are $[4/15], [9/20], [9/20]$ for $T=0.1, 0.01, 0.001$, respectively, which implies it needs more moments to simulate systems at low temperature than high temperature. Similarly, for TD-DMRG, the time correlation functions usually need longer time evolution to decay to zero at low temperature.

\begin{figure}[htbp]
    \centering
    \includegraphics[width=1.0\linewidth]{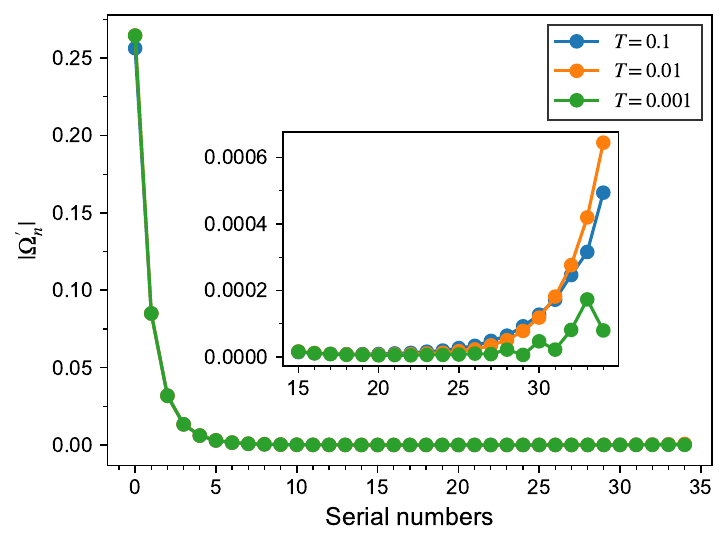}
    \caption{\label{fig-5}{The absolute value of rescaled moments $\left|\Omega_n^\prime\right|$ of the Hubbard model. These rescaled moments are used to calculate the correlation functions Eq.~(\ref{eqn-corr-ca}), so here $A=c_1^\dagger$, $B=c_1$ in Eq.~(\ref{eqn-3}).}}
\end{figure}

In practical calculations, we rescale the Hamiltonian by a factor $r<1$ to suppress exponential growth in the values of moments. Using the rescaled Hamiltonian $H^\prime=rH$ to evaluate the moments in Eq.~(\ref{eqn-3}), the rescaled moments $\Omega_n^\prime$ relate to the original moments $\Omega_n$ as $\Omega_n^\prime=r^{n}\Omega_n$. Consequently, the time variable transforms into $t^\prime=r^{-1}t$. Since the memory kernel $K_{n}(t)$ is derived from Eqs.~(\ref{eqn-13}) and (\ref{eqn-14}), we reformulate $K_n(t)$ explicitly in terms of its dependencies: $K_n\left(t;\{\Omega_m\}\right)$, and it is easy to realize $K_n\left(t;\{\Omega_m\}\right)=K_n\left(t^\prime;\{\Omega^\prime_m\}\right)$. Fig.~\ref{fig-5} presents the absolute values of rescaled moments $\left|\Omega_n^\prime\right|$ of the Hubbard model used to calculate the correlation functions Eq.~(\ref{eqn-corr-ca}), where we set $r=0.5$.

\begin{figure}[htbp]
    \centering
    \includegraphics[width=1.0\linewidth]{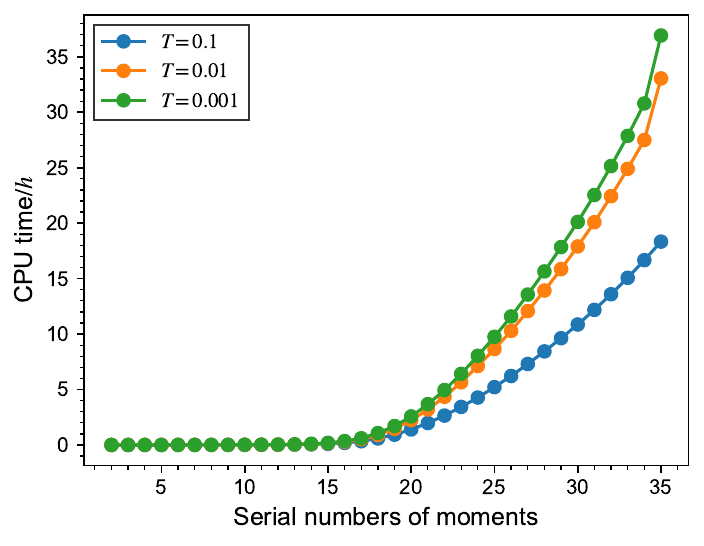}
    \caption{\label{fig-6}{The cumulative computation CPU time after evaluating the $n$-th rescaled moment $\Omega_n^\prime=\left(\mathcal{L}^nc_1^\dagger,c_1\right)/(c_1^\dagger c_1)$ of the Hubbard model.}}
\end{figure}

Finally, we compare the computational efficiency between MKCT-DMRG and TD-DMRG. As the dominant costs arise from moment calculations (other steps are negligible), Fig.~\ref{fig-6} plots the cumulative computation CPU time after evaluating the $n$-th moment. The polynomial growth of time with $n$-th moment stems from increasing entanglement entropy in the operator $\mathcal{L}^nO$, which necessitates more iterative steps to achieve moment convergence at higher orders. This accelerated cost growth implies that MKCT-DMRG may become impractical when moment convergence requires prohibitively long computations. Crucially, Tab.~\ref{tab-1} demonstrates that MKCT-DMRG reduces the CPU time for correlation function calculations obviously compared with TD-DMRG, confirming its significant efficiency advantage.

\begin{table}[htbp]
  \centering
  \caption{CPU time required for correlation function calculations (h)}
  \label{tab-1}
    \begin{tabular}{l |@{\hspace{0.5cm}} c @{\hspace{0.5cm}} c @{\hspace{0.5cm}} c}
      \hline
       & $T=0.1$ & $T=0.01$ & $T=0.001$ \\
      \hline
      MKCT-DMRG & 1.98 & 20.09 & 22.54 \\
      TD-DMRG & 36.33 & 75.56 & 76.06 \\
      \hline
    \end{tabular}
\end{table}

\subsection{Electronic friction} \label{sec:sec3B}

To further verify the effectiveness of our approach, we now compute the electronic friction using our MKCT-DMRG method and benchmark the results against TD-DMRG. The electronic friction\cite{PhysRevB.96.104305,PhysRevB.97.064303} represents the first order correction to the Born-Oppenheimer approximation, providing a fundamental mechanism for understanding nonadiabaticity in a metallic bath. To overcome the limitation of computational cost in nonadiabatic dynamics, the generalized Langevin dynamics framework has emerged as a practical computational paradigm for simulating nonadiabatic dynamics at the molecule-metal interfaces, thereby capturing essential electron transfer between the molecule and the metal surfaces while maintaining computational feasibility. The key to simulate Langevin dynamics is calculating the electronic friction correctly.

The following Hubbard-Holstein model offers an excellent platform for investigating strongly correlated systems featuring electron-phonon (el-ph) coupling, 
\begin{align}
H=&H_\text{Hub} + H_\text{osc} + H_\text{int},  \\
H_\text{Hub}=&  \sum_{i,\sigma} \epsilon_i n_{i,\sigma} + t\sum_{i,\sigma}\left( c_{i,\sigma}^\dagger c_{i+1,\sigma}+ \text{h.c.} \right) \nonumber \\
               &+ U\sum_i n_{i\uparrow }n_{i\downarrow },    \\
H_\text{osc}=& \frac{p^2}{2m} + \frac{1}{2}\omega^2x^2,    \\
H_\text{int}=& \sqrt{2}gx n_1,
\end{align}
where $\epsilon_1=E_d$, indicate an impurity site, and $\epsilon_i=\epsilon$ for $i>1$. This is a quantum-classical hybrid model, in which the electrons are regarded as quantum, and the lattice vibration of the leftmost impurity site is regarded as a classical harmonic oscillator. The whole electronic Hamiltonian is 
\begin{align}
    H_\text{el}=&E(x)\sum_\sigma n_{1\sigma} +\epsilon \sum_{i\ne 1,\sigma}n_{1\sigma} +t\sum_{i,\sigma}\left(c_{i\sigma}^\dagger c_{i+1,\sigma} + \text{h.c.}\right) \nonumber \\
    &+ U\sum_i n_{i\uparrow}n_{i\downarrow},
\end{align}
where $E(x)=E_d + \sqrt{2}gx$.

For this model, at equilibrium, the electronic friction can be obtained by\cite{PhysRevB.96.104305}
\begin{equation}
\gamma(x)=-\pi \hbar \left[\partial_{x} E(x)\right]^2 \sum_\sigma \int \left[P(x,\epsilon)\right]^2 \partial_{\epsilon} f(\epsilon) \mathrm{d} \epsilon,
\end{equation}
where $f(\epsilon)$ is the Fermi function, $f\left(\epsilon\right)=\left(\mathrm{e}^{\beta\epsilon}+1\right)^{-1}$, $P(x,\epsilon)\equiv -\frac{1}{\pi}\operatorname{Im}G(x,\epsilon)^{R}$. The retarded Green's function in time domain is defined as
\begin{equation}
G^R_{i,j}\left(t_1,t_2\right)=-\frac{\mathrm{i}}{\hbar}\theta\left(t_1-t_2\right)\left\langle\left\{\hat{c}_i\left(t_1\right),\hat{c}_j^\dagger\left(t_2\right)\right\}\right\rangle.
\end{equation}

\begin{figure}[htbp]
    \centering
    \includegraphics[width=1.0\linewidth]{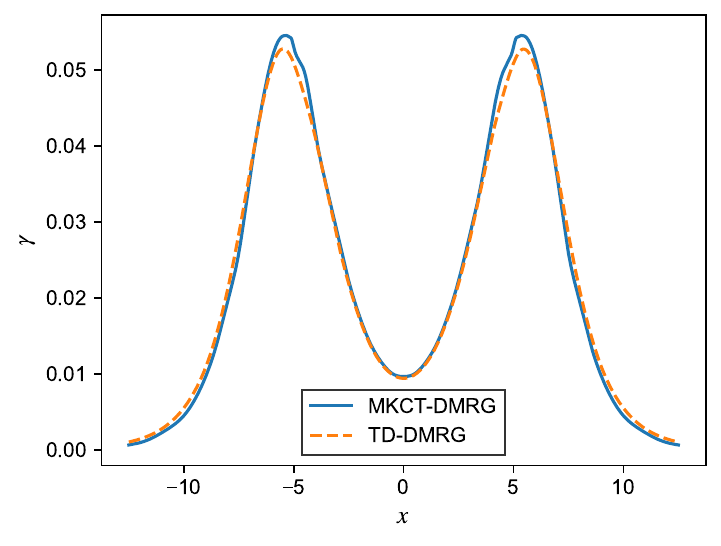}
    \caption{\label{fig-7}{Electronic friction according to MKCT-DMRG and TD-DMRG calculations at temperature T = 0.1.  The parameters are $E_d = \epsilon=-0.5$, $g = 0.075$, $t = 0.3$, $U = 1.0$, the Pad\'e orders is $[M_1/M_2]=[10/25]$, and we set $\hbar = 1$.}}
\end{figure}

As shown in Fig.~\ref{fig-7}, the electronic friction of the Hubbard-Holstein model at $T=0.1$ with 20 electronic sites, calculated using both our MKCT-DMRG method and the benchmark TD-DMRG method, exhibits excellent quantitative agreement, confirming the validity of the MKCT-DMRG approach.

\section{Conclusions} \label{sec:sec4}

In summary, we have developed MKCT-DMRG, a novel framework combining memory kernel coupling theory (MKCT) with density matrix renormalization group (DMRG) for accurately simulating dynamical correlation functions in strongly correlated, quasi-one-dimensional quantum systems. Successful application to representative models demonstrates that MKCT-DMRG yields results in excellent agreement with established time-dependent DMRG (TD-DMRG) benchmarks. Meanwhile, MKCT-DMRG spends less computational costs than TD-DMRG in our calculations. This establishes MKCT-DMRG as a powerful and promising tool for tackling challenging dynamical properties in strongly correlated quantum systems.

Looking forward, while the Padé approximant employed for the memory kernels proves effective, the empirical selection of its orders ($M_1$ and $M_2$) represents a current limitation. Future work should therefore focus on developing a systematic approach to approaching the memory kernels, thereby enhancing the robustness and broader applicability of the MKCT-DMRG method.

\begin{acknowledgments}
W.D. thanks the funding from National Natural Science Foundation of China (No. 22361142829) and Zhejiang Provincial Natural Science Foundation (No. XHD24B0301). Y.L. thanks Wei Liu and Ruihao Bi for helpful suggestions on MKCT. 
\end{acknowledgments}

\bibliography{mkct_dmrg}

\end{document}